\documentclass[aps,twocolumn,pra,superscriptaddress]{revtex4-2}

\usepackage{tikz}
\usepackage{amssymb} \usepackage{color,graphicx} \usepackage{amsmath}
\usepackage{amsbsy} \usepackage{amsthm}
\usepackage{svg}
\usepackage{float} \usepackage{braket}
\usepackage{placeins}
\usepackage[colorlinks=true,citecolor=blue,linkcolor=red,urlcolor=red]{hyperref}
\usepackage[sort&compress]{natbib}
\usepackage{comment}
\usepackage{stackengine}
\usepackage{soul}
\usepackage{siunitx}

\newcommand{\bq}{\begin{equation}} \newcommand{\eq}{\end{equation}}
\newcommand{\bqali}{\bq\begin{aligned}}
\newcommand{\eqali}{\end{aligned}\eq}
\newcommand\D{\operatorname{d}\!}

\renewcommand\k{{\bf k}}
\newcommand\p{{\bf p}}

\newcommand\z{{\bf z}}

\newcommand\x{{\bf x}}

\newcommand\com[2]{[#1,#2]}

\newcommand\rC{r_\text{\tiny C}}

\begin{document}

\author{Davide Giordano Ario Altamura}
\email{davidegiordanoario.altamura@phd.units.it}

\affiliation{Department of Physics, University of Trieste, Strada Costiera 11, 34151 Trieste, Italy}
\affiliation{Istituto Nazionale di Fisica Nucleare, Trieste Section, Via Valerio 2, 34127 Trieste, Italy}

\author{Andrea Vinante}
\affiliation{CNR-IFN, I-38123, Trento, Italy}
\affiliation{Fondazione Bruno Kessler (FBK), I-38123, Trento, Italy}

\author{Matteo Carlesso}
\affiliation{Department of Physics, University of Trieste, Strada Costiera 11, 34151 Trieste, Italy}
\affiliation{Istituto Nazionale di Fisica Nucleare, Trieste Section, Via Valerio 2, 34127 Trieste, Italy}

\title{
Improved bounds on collapse models from rotational noise of LISA Pathfinder}

\date{\today}
\begin{abstract}  

{
{We study spontaneous wavefunction collapse models, which modify the Schrödinger equation with nonlinear and stochastic terms offering a solution to the quantum measurement problem. We find that constraints on the phenomenological parameters of the Continuous Spontaneous Localization (CSL) model can be tightened using the recent analysis of LISA Pathfinder’s angular motion data.
We derived a stronger constraint on the CSL model than the previously achieved with translational motion.
Additionally, we identify general conditions under which rotational measurements provide an advantage over translational ones in testing collapse models.
These results enhance the experimental strategies for probing fundamental modifications of quantum mechanics.}
}
\end{abstract}
\maketitle
\section{Introduction}

Spontaneous wavefunction collapse models \cite{bassi2003dynamical,bassi2013models,carlesso2022present} represent a possible answer to the well-known  measurement problem in quantum mechanics.
They modify the Schr\"odinger equation by adding non-linear, {mass-proportional} and stochastic terms: while microscopic systems are essentially not affected, the effect of these {terms}  become increasingly stronger for larger systems, {resulting} in the suppression of macroscopic superposition states.
Among these models, the most studied is the Continuous Spontaneous Localisation (CSL) model \cite{ghirardi1990markov,pearle1976reduction,pearle1989combining}. It is characterised by two phenomenological constants: the collapse rate $\lambda$ and the spatial resolution of the collapse $\rC$.
Two main theoretical predictions  {exist} in literature for these constants, the {original}  one {proposed by Ghirardi, Rimini and Weber \cite{ghirardi1986unified}}
$\lambda=10^{-16}\,\mathrm{s^{-1}}$ and $\rC=10^{-7}\,\mathrm{m}$, 
and that proposed by Adler \cite{adler2007lower}, who proposed the ranges $\lambda\sim10^{-8\pm 2}\,\mathrm{s^{-1}} $ for $\rC=10^{-7}$\,m and $\lambda=10^{-6\pm 2}\,\mathrm{s^{-1}} $ for $\rC=10^{-6}$\,m.
{It must be noted  that these parameters are phenomenological, so their values should be eventually constrained by experiments}
\cite{carlesso2022present}. Currently, the strongest bounds on the CSL parameters come from the non-interferometric class of experiments \cite{bahrami2014proposal,nimmrichter2014optomechanical,diosi2015testing, carlesso2022present}. {Such tests are based on detecting the Brownian-like motion induced by the collapse dynamics \cite{donadi2023collapse}, and as such they do not require quantum superpositions.}

{Among these experiments, gravitational wave detectors provide the strongest constraint on the CSL parameters in the region of large $\rC$.
In particular, the strongest bound has been inferred from the space mission LISA Pathfinder \cite{carlesso2016experimental,carlesso2018non}, the technological demonstrator of the future space-based gravitational wave detector LISA.
In a nutshell, LISA Pathfinder consists of two test masses in nominal free-fall inside the same satelite, with the main goal of testing the accuracy of free-fall geodesic motion at level of $\sim 10^{-15}$ m/s$^2/\sqrt{\mathrm{Hz}}$. To this end, the distance between the two test masses is measured with high accuracy by an interferometric setup. Besides tracking the relative translational motion, the experiment is also able to monitor the relative angular motion between the two test masses. Angular acceleration data have been analyzed in detail in a recently published article \cite{armano2024nano}. 

In this work, we employ the latter data to infer a new bound on the CSL parameters, which is stronger compared to that previously inferred from translational motion \cite{carlesso2016experimental}.
We identify the origin of the stronger bound on CSL from rotational motion and we formulate a criterion to determine which  between  degree of freedom, translational or rotational, provides the stronger bound, depending on different experimental conditions.}

\section{The model}
By suitably averaging the non-linear and stochastic dynamics of the wavefunction, one obtains the corresponding master equation of the CSL model for the statistical operator $\hat \rho(t)$. Such a master equation reads \cite{bassi2003dynamical}
\begin{equation}\label{eq.master}
\frac{\D \hat\rho(t)}{\D t}=-\frac{i}{\hbar}\left[\hat{H},\hat{\rho}(t)\right]+\mathcal D[\hat \rho(t)],    
\end{equation}
 with $\hat H$ indicating the quantum mechanical Hamiltonian, and
\begin{equation}
    \mathcal D[\hat \rho(t)]=-\frac{\lambda \rC^3}{2\pi^{3 / 2} m_0^2} \int \D \k\, e^{-k^2 \rC^2}\com{\hat{\mu}(\k)}{\com{\hat{\mu}(-\k)}{ \hat{\rho}(t)}},
    \label{Lindblad}
\end{equation}
where $m_0$ is a reference mass chosen equal to the mass of a nucleon and $\hat{\mu}(\k)=\int \D\z \,e^{i\k \z} \sum_n m_n \delta(\z-\hat{\x}_n)$ is the Fourier transform of the mass density operator, $m_n$ and $\hat{\x}_n$ are respectively  the mass and the position operator of the $n$-th particle of the system. 
The CSL effects on the dynamics of the observables of interest can be obtained by considering an unravelling of Eq.~\eqref{eq.master}, which consists of a standard Schr\"odinger equation with an additional stochastic potential of the form \cite{fu1997spontaneous,carlesso2016experimental}
\begin{equation} 
\hat{V}_{\text{\tiny CSL}}(t)=-\frac{\hbar  \rC^{3/2} \sqrt{\lambda} }{\pi^{3 / 4} m_0^2} \int \D \k\, e^{-\frac{k^2\rC^2}{2}} \hat{\mu}(\k) w(\k, t),
\end{equation}
where $w(\k, t)$ is a white, Gaussian noise, which is fully characterised by its mean $\mathbb{E}[w(\k, t)]=0$ and its two-point correlation $\mathbb{E}[w(\k, t) w(\p, s)]= (2\pi)^{3}\delta(t-s) \delta^{(3)}(\k+\p)$. 
Such a stochastic potential acts on the $n$-th particles of the system as a stochastic force that can be computed as
\begin{equation}\label{Fn}
    \hat{\bf{F}}_n=\frac{i}{\hbar}[\hat V_{\text{\tiny CSL}},\hat{\bf {p}}_n].
\end{equation}
When considering the motion of the center of mass of a rigid body, one sums over the particles belonging to it and obtains the total force, which reads
\begin{equation}
    \hat{{\bf F}}(t)=\frac{\hbar\rC^{3 / 2} \sqrt{\lambda}}{ \pi^{3/4}  m_0} \int  \D \k \, \hat{\mu} (\k)e^{-\frac{\k^2 \rC^2}{2}}\k w(\k, t).
\end{equation}
In the particular case of LISA Pathfinder, where there are two test masses, the monitored degrees of freedom are the those of the relative motion of one mass with respect to the other. These can be effectively accounted by the one-dimensional relative distance $\hat x_\text{rel}$ and corresponding momentum that are computed along the axis connecting the two centers of mass. The Langevin equations describing the relative motion are given by
\begin{equation}
    \begin{aligned}
\frac{\D}{\D t} \hat{x}_{\text {rel}}(t) & =\frac{2 \hat{p}_{\mathrm{rel}}(t)}{m}, \\
\frac{\D}{\D t} \hat{p}_{\text {rel}}(t) & =-\frac{m}{2} \omega_0^2 \hat{x}_{\text {rel}}(t)-\gamma \hat{p}_{\text {rel}}(t)+\hat F_{\text {rel}}(t),
\end{aligned}
\end{equation}
where $\hat F_{\text{rel}}=\frac{1}{2}(\hat F_1(t)-\hat F_2(t))$, where $\hat F_i$ is the stochastic force acting on the $i$-th mass along the aforementioned axis.
From these equations, it is possible to compute the force density noise spectrum (DNS) reading $S^{F}_\text{\tiny CSL}(\omega)=\frac{1}{2\pi}\int\D\Omega\,\mathbb{E}\left[\braket{ \hat F_{\text{rel}}\left(\omega\right)\hat F_{\text{rel}}\left(\Omega\right)}\right]$, where $\hat F_\text{rel}(\omega)$ is the Fourier transform of $\hat F_\text{rel}(t)$. The theoretical expression of $S^F_\text{\tiny CSL}(\omega)$ is then compared with the experimental one $S^F_\text{exp}(\omega)$ to derive experimental bounds on the CSL parameters, namely $S^F_\text{\tiny CSL}(\omega)\leq S^F_\text{exp}(\omega)$.

\begin{figure*}[t]
    \centering
    \includegraphics[width=0.8\linewidth]{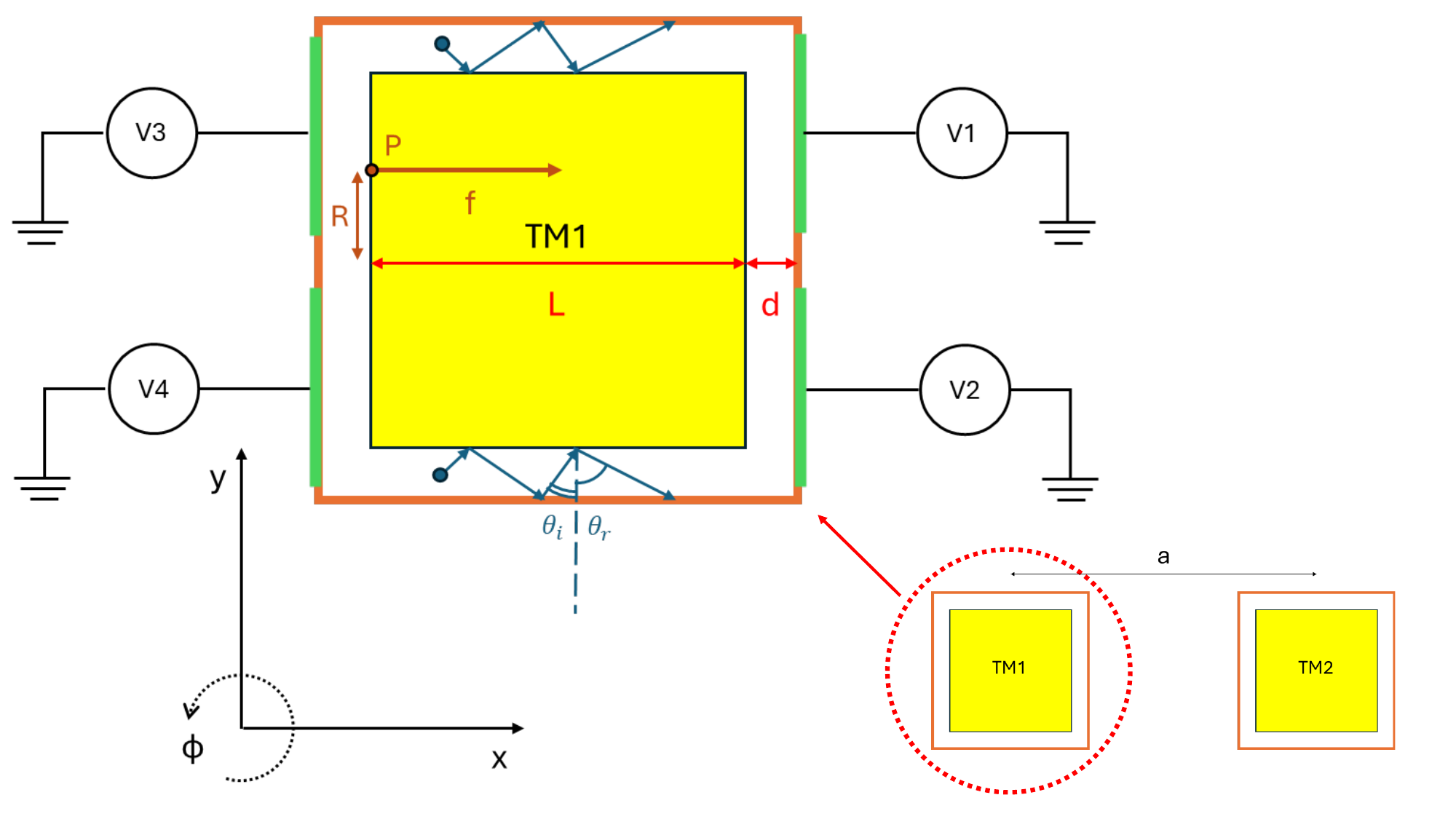}
    \caption{Schematic of the experimental setup. Bottom right: disposition of the two test masses (TM) separated by a distance $a$. Top left: the single cubic TM, of length $L$, enclosed in a box comprising control electrodes with a gap $d$.
    The total noise force due to the actuators, which are depicted in green, is represented as the brown vector $f$ applied at the point $P$ to the test mass.
    {This generates an additional torque noise on the system.}
    {The blue particles represent the residual gas impinging on the test mass with incidence angle $\theta_i$ and reflection angle $\theta_r$.}} 
    \label{LISA}
\end{figure*}

In a similar way,  we can write a set of analogous equations for the rotational motion, which read
\begin{equation}
    \begin{aligned}
\frac{\D}{\D t} \hat{\phi}_{\text {rel}}(t) & =\frac{2 \hat{L}_{\mathrm{rel}}(t)}{m}, \\
\frac{\D}{\D t} \hat{L}_{\text {rel}}(t) & =-\frac{I}{2} \omega_0^2 \hat{\phi}_{\text {rel}}(t)-\gamma \hat{L}_{\text {rel}}(t)+\hat \tau_{\text {rel}}(t),
\end{aligned}
\end{equation}
where $\hat\phi_{\text{rel}}=\hat\phi_1-\hat\phi_2$ is the relative orientational angle of the two masses, $\hat L_{\text{rel}}=\frac{1}{2}(\hat L_1-\hat L_2)$ the corresponding angular momentum, and  $\hat \tau_{\text{rel}}=\hat \tau_1-\hat \tau_2$ the relative CSL-induced stochastic torque. 
{The expression of the torque $\hat{\tau}_i$ along the axis perpendicular to the  $x$-$y$ plane (see Fig.~\ref{LISA}) for each of the two test masses reads:
\begin{equation} 
\hat\tau_i=\frac{\hbar\rC^{3 / 2} \sqrt{\lambda}}{ \pi^{3/4}  m_0} \int  \D \k \int\D \z\, \hat{\mu}_i (\z)e^{-\frac{\k^2 \rC^2}{2}-i \k\z} w(\z, t)\,(\z\times\k)_\perp,
\end{equation} 
where $(\z\times\k)_\perp$ is the component of the vector along the axis perpendicular to the $x$-$y$ plane.} 
Given the above expression,  the relative torque
has zero mean $\mathbb{E}\left[\hat \tau_{\text{rel}}(t)\right]=0$ and correlation function reading $\mathbb{E}\left[\hat \tau_{\text{rel}}(t)\hat \tau_{\text{rel}}(s)\right]=\hbar^2 \eta_R\delta(t-s)$, where 
 $\eta_{\mathrm{R}}$ is the rotational diffusion constant due to CSL. Notably, this quantity depends on the geometry of the system and the parameters of the CSL model.
Finally, one computes the CSL-induced torque density noise spectrum, which is given by
\begin{equation}
    S^{\tau}_\text{\tiny CSL}(\omega)=\frac{1}{2\pi}\int\D\Omega\,\mathbb{E}\left[\braket{ \hat \tau_{\text{rel}}\left(\omega\right)\hat \tau_{\text{rel}}\left(\Omega\right)}\right].
\end{equation}
Due to the form of correlation of the CSL-induced force and torque noises, one finds that $S^{F/\tau}_\text{\tiny CSL}=\hbar^2 \eta_{\mathrm{V}/\mathrm{R}}$, where the general expression for $\eta_i$ is given in Appendix \ref{A}.

\section{The experimental bound}

The two test masses of LISA Pathfinder are solid cubes of gold-platinum alloy with side $L=4.6\, \mathrm{cm}$ and mass $m=1.928\, \mathrm{kg}$.
To place a conservative bound to the CSL parameters, we attribute all the torque noise measured in the experiment {$S^\tau_{\text{exp}}(\omega)$} to the CSL collapse noise, {namely we impose $S^\tau_{\text{\tiny{CSL}}}(\omega)\le S^\tau_{\text{exp}}(\omega)$}.
{This condition translates into an upper bound on the collapse rate $\lambda$ as a function of the length parameter $\rC$.}
Since the output of the experiment is expressed as a relative angular acceleration noise spectrum $S_{\Delta \gamma}$, we have to transform it to the relative torque noise
spectral density. This can be done employing the relation  \cite{carlesso2016experimental}:
\begin{equation}
S^\tau_{\text{exp}}=\frac{1}{4}I^2 S_{\Delta \gamma} ,
\end{equation}
where $I=mL^2/6$.
Concretely, we place the bound by inferring the {minimum} value of the experimentally measured torque DNS {from Fig.~11 of Ref.~\cite{armano2024nano} in which the angular acceleration spectrum $S_{\Delta \gamma}^{1/2}$ is represented.}  
We find the minimum $S^\tau_{\text{exp}}=5.7\times 10^{-34} \,\mathrm{N^2}\, \mathrm{m^2}/\mathrm{Hz}$ {at the frequency $3\times10^{-3}\;\mathrm{Hz}$. 
\begin{figure}[t]
    \centering
    \includegraphics[width=\linewidth]{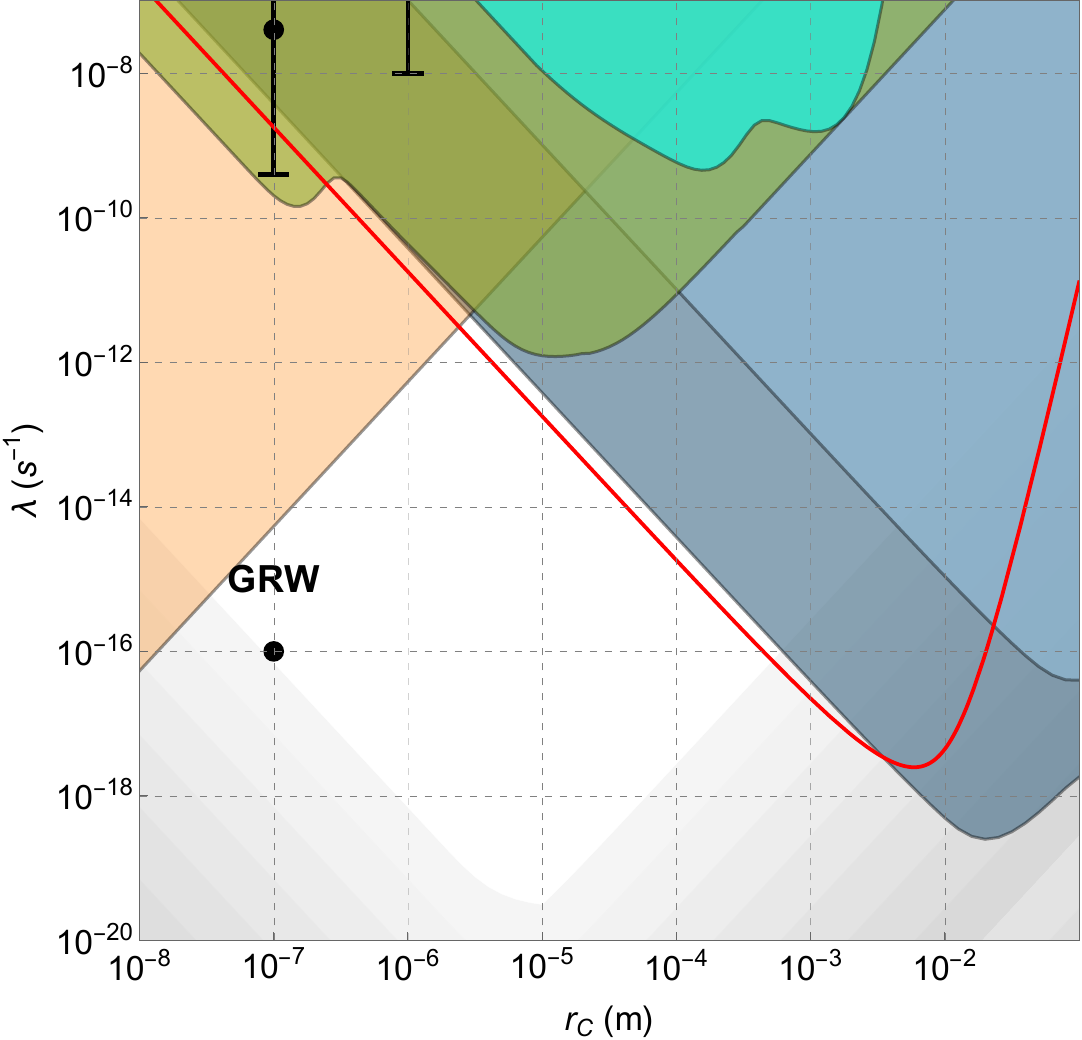}
    \caption{Exclusion plot for the CSL parameters $\lambda$ and $\rC$. 
    We present a comparison between the red curve, which represents the rotational bound from LISA Pathfinder obtained in this work, with the existing bounds present in literature. 
    The cyan area represent the region excluded by the another rotational optomechanical experiment \cite{PhysRevA.109.062212}. The green area is excluded by cantilever experiment with multilayer structure \cite{carlesso2018multilayer}.
    The blue areas represent the bound obtained from the gravitational wave detectors\cite{carlesso2016experimental,carlesso2018non,helou2017lisa}: in light blue the one from LIGO, in dark blue is the one derived from the translational spectrum of LISA Pathfinder. The orange area is derived from the spontaneous $\mathrm{X}$-ray emission test \textsc{Majorana} \cite{arnquist2022search}. The gray region is theoretically excluded by requiring that macroscopic superposition should not persist in time \cite{torovs2017colored}. The white area is yet to be explored.}
    \label{BoundLisaRot}
\end{figure}
The corresponding, conservative bound on the CSL parameters is shown in Fig.~\ref{BoundLisaRot} and represents around a factor $2$ improvement with respect to the previous bounds in the region of $\rC$ between $\sim10^{-5.5}$\,m and $\sim10^{-3.5}$\,m 
{\cite{carlesso2016experimental}}.

\section{Discussion}
{We will now discuss why the rotational noise leads to a stronger bound on the CSL model, as apparent from Fig.~\ref{BoundLisaRot}. To this end, we analyze the ratio $\alpha=S^\tau/S^F$ between torque $S^\tau$ and force $S^F$ DNS. This ratio depends on the specific nature of the noise source and how this acts on translational or rotational modes. In the specific case of LISA Pathfinder, the force/torque noise around the minimum is likely dominated by gas collisions. This interpretation is supported by the fact that the noise level is steadily decreasing with time, in agreement with an outgassing mechanism \cite{armano2016sub}. On the other hand, the $\alpha$ ratio may take different values for the CSL noise, depending on the geometry of the test mass.
Since the conservative bound on the CSL parameter space is inferred from the condition $S_{\text{\tiny{CSL}}}/S_{\text{exp}}\le 1$, it is easy to check that if $\alpha_{\text{\tiny{CSL}}} > \alpha_{\text{exp}}$ then the bound on the CSL model from rotational noise will be stronger than the bound from translational noise, and vice versa.
}
 
{Let us first calculate $\alpha_\text{\tiny{CSL}}$ for the single cubic mass of LISA Pathfinder.} The torque and force DNS are given by $S_\text{\tiny{CSL}}^\tau=\hbar^2 \eta^\text{(cube)}_R$ and $S_\text{\tiny{CSL}}^F=\hbar^2 \eta^\text{(cube)}_V$, with $\eta_i^\text{(cube)}$ given in Appendix \ref{A}.
Defining the ratio $\beta = L/\rC$, where $L$ is the length of the cube, and using the expressions for $\eta_i^\text{(cube)}$ in Appendix \ref{A}, one gets the following expression for noise ratio:
\begin{figure}[th]
    \centering
    \includegraphics[width=0.9\linewidth]{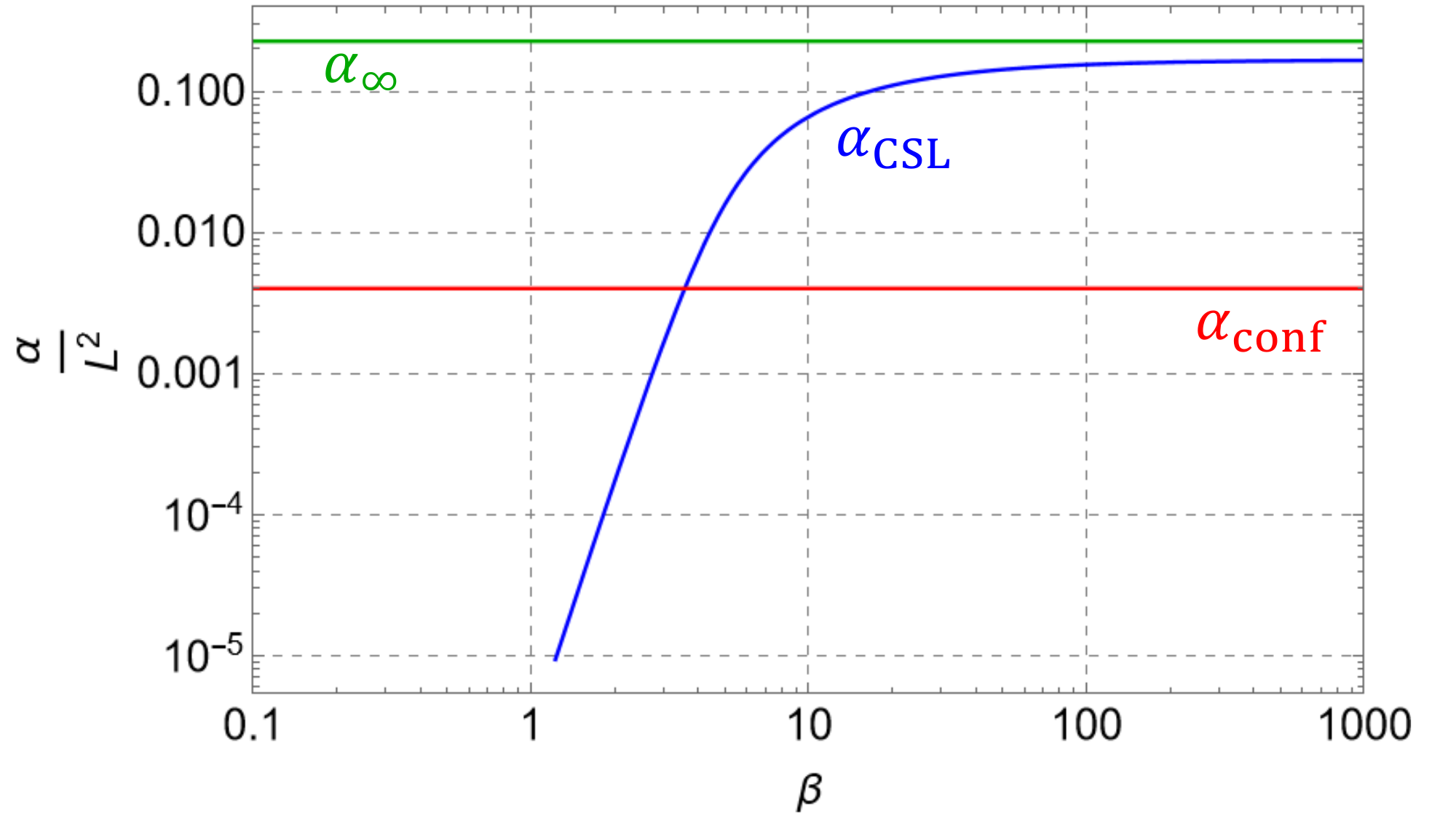}
    \caption{
    Dependence of 
    the ratio $\alpha/L^2=(S^{\tau}/S^{F})/L^2$ on $\beta=L/\rC$.
    The blue line represent the ratio relative to the CSL noise. The red and green lines represent the analogous quantity for the physical thermal noises, respectively in the case of an infinite volume of gas ($\alpha_\infty$) and in the case of a confined gas in an enclosure ($\alpha_{\text{conf}}$).  }
    \label{alpha(beta)}
\end{figure}

\begin{widetext}
    \begin{equation}
        \begin{aligned}
            \frac{\alpha}{L^2}=\frac{-2 \left(\beta ^2+16\right)+e^{\frac{\beta ^2}{4}} \left[8 \left(\beta ^2+8\right)-  \left(\beta ^2+24\right)g(\beta)\right]+e^{\frac{\beta ^2}{2}} \left[-6 \beta ^2-3 g(\beta)^2+\left(\beta ^2+24\right) g(\beta)-32\right]}{6 \left(e^{\frac{\beta^2}{4}}-1\right) \beta ^2 \left[e^{\frac{\beta ^2}{4}} \left(g(\beta)-2\right)+2\right]}
        \end{aligned},
    \end{equation}
\end{widetext}
where $g(\beta)=\sqrt{\pi}\beta\operatorname{erf}(\beta/2)$. 
The relevant range in the case of LISA Pathfinder is that of
 CSL parameter $\rC$ being much smaller than the size of the cube $L$, i.e.~$\beta\gg1$. In such a case, $\alpha$ simplifies to
\begin{equation}
\alpha_{\text{\tiny{CSL}}}\simeq\frac{L^2}{6}.
\end{equation}

\begin{figure}[h]
    \centering
    \includegraphics[width=0.8\linewidth]{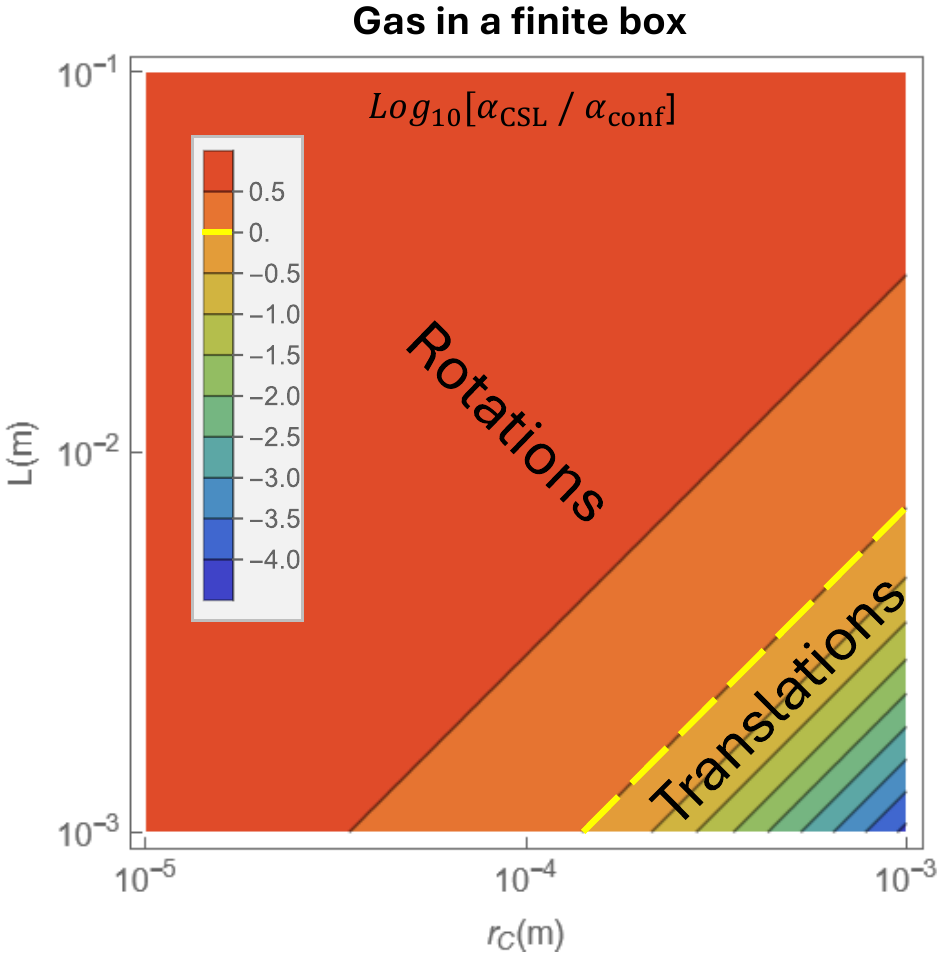}\\
    \includegraphics[width=0.8\linewidth]{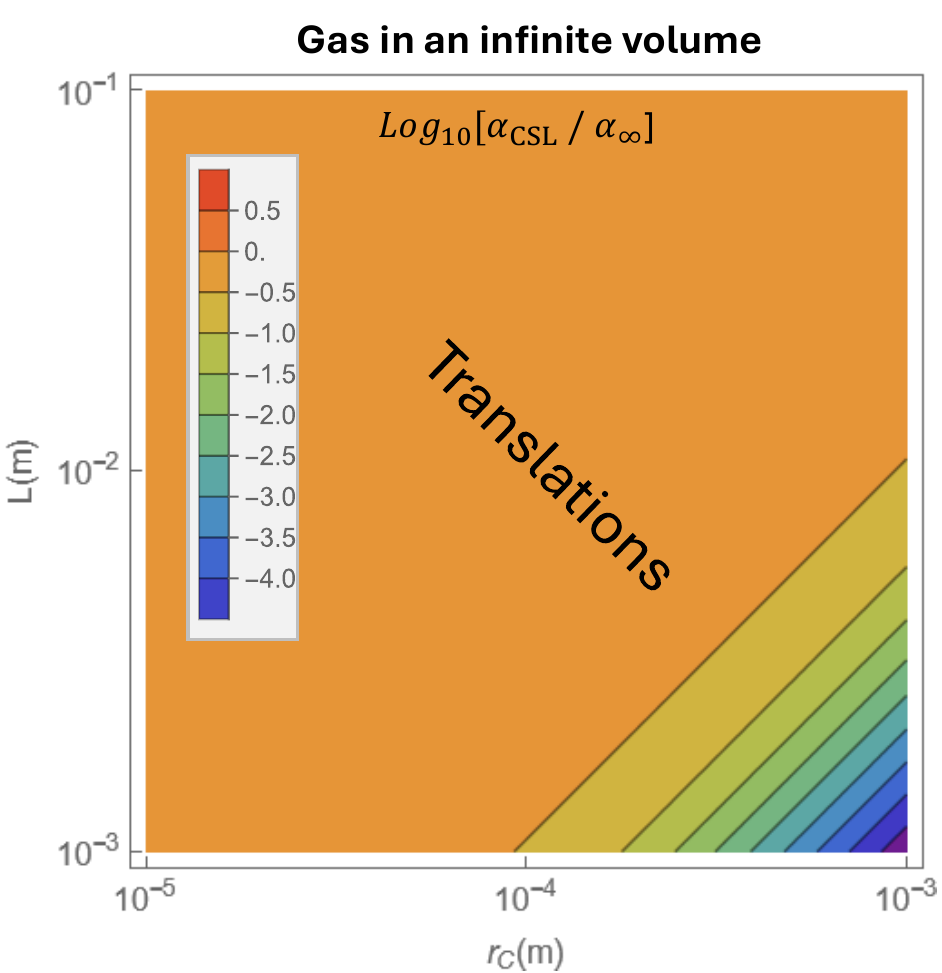}
    \caption{Ratio $\alpha_\text{\tiny{CSL}}/\alpha_{\text{exp}}$ as a function of $\rC$ and $L$. The top panel represents the case in which the setup is enclosed in a box such that of Fig.~\ref{LISA} with $d\ll L$. The rotational dynamics poses stronger bounds on the parameters in the region in which $\operatorname{Log}_{10}[\alpha_\text{\tiny{CSL}}/\alpha_{\text{conf}}]$ is positive (red and dark orange region). 
    The bottom panel represents the case in which the setup is surrounded by an infinite volume of gas. In this case, the translational dynamics is preferable for any choice of $\rC$ and $L$. }
    \label{ContourPlot}
\end{figure}

{Let us estimate the $\alpha$ ratio for the noise from residual gas, which is believed to be the main source of physical noise in  LISA Pathfinder.} This noise arises from the inelastic collisions of the gas particles with the cubic test masses. 
For a cube in a infinite volume of gas, it is found \cite{cavalleri2010gas} that the ratio of rotational
to {translational} noise is $\alpha_\infty=0.226 L^2>\alpha_{\text{\tiny{CSL}}}$.
Thus, in this particular case, the bound on the CSL model from the rotational noise is expected to be weaker than the bound from translational noise.

However, this is not the case relevant to LISA Pathfinder. In fact, the test masses in the actual experimental setup are surrounded by an enclosure, the gravitational reference sensor (GRS), comprising the electrodes for inertial sensing, which is slightly larger than the test mass (the gap $d$ is roughly one tenth of $L$, see Fig.~\ref{LISA}). In this case the correlation between consecutive collisions of each gas particle with the test masses leads to an increase of both damping and noise \cite{cavalleri2009increased}. The relative increase of {translational} noise is however larger than the increase of the rotational noise, leading to a lower value of $\alpha$.
From Fig.~5 of Ref.~\cite{cavalleri2009increased} it is possible to infer the new value of the ratio for this particular configuration $\alpha_{\text{conf}}\simeq0.04 L^2<\alpha_{\text{\tiny{CSL}}}$.
We can thus conclude that for this more relevant case of tight gas confinement, rotational noise should provide a stronger bound on the CSL model, which is in agreement with Fig. \ref{BoundLisaRot}.

{Fig.~\ref{alpha(beta)} represents a direct comparison of the ratio $\alpha/L^2$ as a function of $\beta$  for the noise sources considered above. From a comparison of the coefficient for the CSL case ($\alpha_{\text{\tiny{CSL}}}$, blue line) and that of the confined gas ($\alpha_{\text{conf}}$, red line), we can see that for large enough values of $\beta$ the rotational noise becomes favorable for a system enclosed in a box. Conversely, the coefficient for the case of an infinite volume of gas   ($\alpha_\infty$, green line) is larger than the CSL one for all values of $\beta$, meaning that monitoring the translational noise is always favorable.} 

Moreover, in Ref.~\cite{carlesso2018non}, {under the assumption that the residual gas was the dominant source of noise}, the conversion of the force DNS ($S^F_{\text{conv}}=3.15\times 10^{-30}\, \mathrm{N^2}/\mathrm{Hz}$) into the torque DNS $S^\tau_{\text{exp}}=\alpha_{\text{conf}}S^F_{\text{exp}}$ 
was used to place a hypothetical bound on the CSL parameters. Namely, the hypothetical bound is a factor 2 stronger than that obtained above.
However, the measured torque DNS in Fig.~11 of Ref.~\cite{armano2024nano} $S^\tau_\text{exp}=5.66\times10^{-34}\,\mathrm{N^2}\, \mathrm{m^2}/\mathrm{Hz}$ is not compatible with the predicted value of $S^\tau_{\text{conv}}\simeq2.66\times10^{-34}\,\mathrm{N^2}\, \mathrm{m^2}/\mathrm{Hz}$. 
This issue was further commented both in Ref.~\cite{armano2024nano} and Ref.~\cite{armano2024depth}.
This inconsistency can be addressed to the correlation existing between the noises coming from the four electrodes of the GRS.
In fact, an asymmetry in the electrode voltage noise was noticed during the measurements. In particular, the electrodes 2 and 3 where slightly but consistently noisier than electrodes 1 and 4 (see Fig.~\ref{LISA}).

In general, the choice between rotational and translational degrees of freedom is made
    depending on the expected dominant physical source of noise affecting the setup and its geometry. For this purpose, we studied the ratio between $\alpha_{\text{\tiny{CSL}}}$ and $\alpha_{\text{exp}}$ for different noises and sizes of the setup. This is shown in Fig.~\ref{ContourPlot}. In the case of a setup enclosed in a finite box, such as that of LISA Pathfinder (cf.~Fig.~\ref{LISA}), with a gap much smaller than the dimension of the mass ($d\ll L$), rotational noise provides a stronger bound than translational one for a large set of values of $L$ and $\rC$ (see top panel).
Conversely, for large or infinite values of the gap $d$ ($d\simeq L$ or $d\gg L$),
translational degrees of freedom are the preferrable choice for any value of $L$ and $\rC$ (see bottom panel).

\section{Conclusions}

In this work, we derived the first experimental bound on the CSL parameters from the rotational noise of LISA Pathfinder. The obtained bound improves that previously derived from the translational noise and becomes the most stringent constraint on the CSL model for the values of $\rC$ between the micro and millimeter scale. The bound is derived  maintaining a conservative approach that accounts for the entirety of the rotational noise of the LISA Pathfinder.
The result strengthens the claim that a dedicated experiment measuring the torque noise of a system with a suitably chosen geometry can, in principle, exclude new regions of the parameter space \cite{PhysRevA.109.062212}.

{When finalising this paper, we became aware of a related work \cite{dai2024updatingconstraintquantumcollapse} that sets a much stronger bound on the CSL model by assuming that the totality of the LISA Pathfinder noise is fully modeled and understood, based on a substantial agreement with an outgassing model. Such a  bound on CSL is thus set based solely on the experimental error bar. While it is plausible to assume that the majority of the observed noise can be indeed attributed to known physical mechanisms, and in particular to gas collisions, the validity of such a procedure in setting a fundamental physics bound is highly questionable. In fact, outgassing is described by very phenomenological and non fundamental laws. More importantly, the relevant parameters for a full and comprehensive modeling and subtraction of gas-related noise, for instance the gas pressure inside the spacecraft, are not independently measured or controlled in the experiment. As a counterexample, the procedure used in Refs.~\cite{vinante2016upper,vinante2017improved,Vinante_2020} can be considered legitimate as in these experiments the thermal noise follows a simple fundamental law (the fluctuation-dissipation relation) and the relevant parameters for an independent estimation (temperature, damping) are directly under the control of the experimentalist.} 

Finally, we comment that implementing the analysis for the LISA Pathfinder translational noise --- with the same conservative approach we applied here --- would recover the bound already derived in Ref.~\cite{carlesso2018non} and shown in dark blue in Fig.~\ref{BoundLisaRot}, as both Refs.~\cite{carlesso2018non,dai2024updatingconstraintquantumcollapse}  employ the same dataset.

\section{Acknowledgments}
We thank W.J.~Weber for drawing our attention on the new analysis of LISA Pathfinder rotational noise. 
We acknowledge support from the QuantERA II Programme (project LEMAQUME) that has received funding from the European Union’s Horizon 2020 research and innovation programme under Grant Agreement No 101017733. We also acknowledge the PNRR PE Italian National Quantum Science and Technology Institute (PE0000023), {the EU EIC Pathfinder project QuCoM (101046973)} and the University of Trieste (Microgrant LR 2/2011).

\newpage

\appendix
\vspace*{\fill}

   \onecolumngrid
   
\section{CSL Diffusion coefficients}
\label{A}
The CSL diffusion coefficients depend on the mass density $\mu(\mathrm{r})$, they read
$$
\begin{aligned}
\eta_{\mathrm{V}} & =\frac{\lambda r_C^3}{\pi^{3 / 2} m_0^2} \int \mathrm{d}^3 \mathbf{k}\, e^{-\rC^2 k^2} k_x^2|\tilde{\mu}(\mathbf{k})|^2, \\
\eta_{\mathrm{R}} & =\frac{\lambda r_C^3}{\pi^{3 / 2} m_0^2} \int \mathrm{d}^3 \mathbf{k}\, e^{-\rC^2 k^2}\left|k_y \partial_{k_z} \tilde{\mu}(\mathbf{k})-k_z \partial_{k_y} \tilde{\mu}(\mathbf{k})\right|^2,
\end{aligned}
$$
where $\tilde{\mu}(\mathbf{k})$ is the Fourier transform of $\mu(\mathbf{r})$.
In the case of a cube, which is the relevant one for our analysis, we have
\begin{equation}
    \eta_{\mathrm{V}}^{\text {(cube) }}=\frac{32 \lambda \rC^4 m^2}{m_0^2 L^6}\left(\frac{\sqrt{\pi}}{2} g\left(\frac{L}{\rC}\right)-1+e^{-\frac{L^2}{4 \rC^2}}\right)^2\left(1-e^{-\frac{L^2}{4 \rC^2}}\right),
\end{equation}
and
\begin{equation}
    \begin{aligned}
& \eta_{\mathrm{R}}^{\text {(cube) }}=\frac{8 \lambda}{3}\left(\frac{m}{m_0}\right)^2\left(\frac{\rC}{L}\right)^6\left(1-e^{-\frac{L^2}{4 \rC^2}}-\frac{\sqrt{\pi}}{2 } g\left(\frac{L}{\rC}\right)\right) \\
& \times\left\{\left(1-e^{-\frac{L^2}{4 \rC^2}}\right)\left[2\left(3-e^{-\frac{L^2}{4 \rC^2}}\right)\left(\frac{L}{\rC}\right)^2+32\left(1-e^{-\frac{L^2}{4 \rC^2}}\right)-\sqrt{\pi}\left[24+\left(\frac{L}{\rC}\right)^2\right] g\left(\frac{L}{\rC}\right)\right]+3 \pi\left( g\left(\frac{L}{\rC}\right)\right)^2\right\},
\end{aligned}
\end{equation}
where $g\left(\frac{L}{\rC}\right)=\sqrt{\pi}\frac{L}{\rC}\operatorname{erf}(\frac{L}{2 \rC})$.

\bibliography{main.bib}

\end{document}